\documentclass[aps,prl,twocolumn]{revtex4}

\usepackage{here}

\usepackage{graphicx}

\newcommand\pictc[5]{\begin{figure}
                       \centerline{
                       \includegraphics[width=#1\columnwidth]{#3}}
                   \protect\caption{\protect\label{fig:#4} #5}
                    \end{figure}            }
\newcommand\pict[4][1.]{\pictc{#1}{!tb}{#2}{#3}{#4}}
\newcommand\rpict[1]{\ref{fig:#1}}

\newcommand\leqt[1]{\protect\label{eq:#1}}
\newcommand\reqtn[1]{\ref{eq:#1}}
\newcommand\reqt[1]{(\reqtn{#1})}

\newcounter{Fig}

\begin{document}
\begin{sloppy}

\title{Multi-gap discrete vector solitons}

\author{Andrey A. Sukhorukov}
\author{Yuri S. Kivshar}

\affiliation{Nonlinear Physics Group, Research School of Physical
Sciences and Engineering, Australian National University,
Canberra, ACT 0200, Australia}
\homepage{http://www.rsphysse.anu.edu.au/nonlinear}

\begin{abstract}
We analyze nonlinear collective effects in periodic systems with
multi-gap transmission spectra such as light in waveguide arrays
or Bose-Einstein condensates in optical lattices. We demonstrate that the inter-band interactions in nonlinear periodic gratings can be efficiently managed by controlling their geometry, and predict novel types of {\em discrete vector solitons} supported by nonlinear coupling between different bandgaps and study their stability.
\end{abstract}

\maketitle

Periodic structures are common in nature, with the crystalline
lattice being the most familiar example. One of the important
common features of such systems is the existence of {\em frequency
gaps} in the transmission spectra which can dramatically affect
both propagation and localization of waves. Moreover, the modern
technology allows creating different structures with an {\em
artificial periodicity}, and the recent examples are {\em photonic
crystals}, which can control propagation and emission of
electromagnetic waves~\cite{Joannopoulos:1995:PhotonicCrystals},
and {\em optical lattices}, which are used to trap and manipulate
atomic Bose-Einstein condensates (BECs)
\cite{Cataliotti:2001-843:SCI}. An unprecedented level of control
over such engineered structures can be realized by tailoring both
location and width of {\em multiple band-gaps} with additional
modulation of the structure parameters. For example, it has been
shown that atomic BEC can demonstrate a rich variety of phase
transitions in optical
superlattices~\cite{Roth:cond-mat/0205412:ARXIV}, and the
reduced-symmetry photonic crystals allow self-localization of
waves in mini-gaps \cite{Mingaleev:2001-5474:PRL}.

The response of many systems becomes {\em nonlinear} at higher
energy densities. This phenomenon may have various physical
origins, such as the charge recombination in biased photorefractive
crystals~\cite{Yeh:1993:IntroductionPhotorefractive}, excitation
of higher energy levels in
semiconductors~\cite{Luther-Davies:2001-19:SpatialOptical}, or
atomic interaction in BEC~\cite{Dalfovo:1999-463:RMP}. In periodic
media, nonlinearity produces a shift of the band-gap spectrum, and
this physical mechanism is responsible for a number of remarkable
effects, including the formation of {\em gap
solitons}~\cite{deSterke:1994-203:ProgressOptics}. Such nonlinear
localized modes can be excited within multiple spectral gaps of a
periodic structure, as was first demonstrated experimentally
for temporal optical pulses in fiber Bragg
gratings~\cite{Eggleton:1996-1223:OL}. Since the pulses extend
over hundreds of grating periods, the gap-soliton dynamics is usually
described by averaged coupled-mode equations with constant
coefficients~\cite{Broderick:1995-5788:PRE}. In contrast, spatial
optical beams in waveguide arrays and matter waves in BEC can span
over a few periods of the structure. Under such conditions the
wave profiles are essentially discretized by the underlying
periodic structure, strongly affecting the properties of {\em
discrete solitons}~\cite{Eisenberg:2002-2938:JOSB,
Sukhorukov:2003-31:IQE}.

Spatial optical solitons associated with the first spectral band
of the multi-band transmission spectrum have been extensively
studied both theoretically and experimentally in arrays of coupled
optical waveguides~\cite{Eisenberg:2002-2938:JOSB}. Very recently,
spatial gap solitons localized in higher-order bands have been
observed as well~\cite{Mandelik:2003-53902:PRL}. Similar
observation of matter-wave gap solitons in BEC loaded into an
optical lattice is expected soon as well. However, the interaction
properties of localized modes {\em which belong to different gaps}
are not known. The existence of gap solitons is directly related
to the band-gap spectrum, and the latter can be fine-tuned in
superlattices. In this Letter, we reveal, for the first time to
our knowledge, the fundamental links between periodic modulation
of the medium parameters and nonlinear wave coupling between
different gaps, and also predict the existence of {\em multi-gap
discrete vector solitons} with nontrivial symmetry and stability
properties. We believe that our results can stimulate and guide
the future experiments on discrete gap solitons in optics and
matter-wave physics.

Self-action and interaction of optical beams in a one-dimensional
periodic structure of coupled optical waveguides with the
normalized refractive index profile $V(x)=V(x+h)$ 
can be described by a set of coupled nonlinear Schr\"odinger equations,
\begin{equation} \leqt{nls}
  i \frac{\partial {\bf \psi}}{\partial z}
  + \frac{\partial^2 {\bf \psi}}{\partial x^2}
  + V(x) \psi
  + {\bf G}(\psi) {\bf \psi}
  = 0,
\end{equation}
where, in the case of optical waveguides, $x$ and $z$ are the
transverse and longitudinal coordinates, respectively, $h$ is the
spatial period, ${\bf \psi} = (\psi^{(1)}, \psi^{(2)}, ...)$ are
the normalized electric field envelopes of several co-propagating
beams having different polarizations or detuned frequencies. It is
assumed that the beams interact incoherently through the Kerr-type
nonlinear change of the refractive index, $G_m = \sum_j \Gamma_{m
j} |\psi^{(j)}|^2$, where $\Gamma_{m=j}$ and $\Gamma_{m\ne j}$ are
the self- and cross-phase modulation coefficients, respectively.

We note that the model \reqt{nls} is equivalent to a system of
the coupled Gross-Pitaevskii equations describing the dynamics of
multi-component BEC in 1D optical lattice, where $\psi^{(m)}$ is
the mean-field wave function for atoms in the $m$-th quantum
state, $z$ stands for time, $V(x)$ is the periodic potential of
an optical lattice, and ${\bf G}$ is the effective mean-field
nonlinearity which appears due to the $s$-wave atom interaction.
Although below we use the terminology from guided wave optics
where the rapid progress in the experimental study of spatial
optical solitons is observed, our results are equally applicable
to the nonlinear dynamics of BEC in optical lattices.

{\em Linear} wave propagation through a periodic structure can be
entirely described by the Floquet-Bloch spectrum of the eigenmode
solutions of Eqs.~\reqt{nls} in the form $\psi = \psi_b \exp({i
\beta z + i K_b x/h})$, where $K_b$ is the normalized Bloch-wave
number and $\beta$ is the propagation constant. The Bloch wave
amplitudes decay exponentially when ${\rm Im}K_b(\beta) \ne 0$,
and this condition defines the location of gaps in the
transmission spectrum. At the gap edges, $K_b = 0,\pi$.

{\em Nonlinearity} manifests itself through an effective change of
the optical refractive index which results in a local shift of the
bands and gaps. This physical mechanism is responsible for the
formation of {\em gap solitons}. When the band shifts are small,
we can seek solutions of Eqs.~\reqt{dnls} near the gap edges
($\beta= \beta_{m}$) in the form of modulated Bloch
waves~\cite{Sipe:1988-132:OL}, $\psi^{(m)} = \varphi^{(m)}
\psi_b^{(m)} \exp(i \beta_{m} z + i K_b^{(m)} x / h)$, and
obtain a system of coupled nonlinear Schr\"odinger equations for
the slowly varying envelopes,
\begin{eqnarray} \leqt{nlsc}
  i \frac{\partial \varphi^{(m)}}{\partial z}
  + \frac{D^{(m)}}{2} \frac{\partial^2 \varphi^{(m)}}{\partial x^2}
  + \sum_j \gamma_{m j} \Gamma_{m j} |\varphi^{(j)}|^2 \varphi^{(m)}
  = 0.
\end{eqnarray}
Here $D^{(m)} = - h^2 \left. \partial^2 \beta / \partial K_b^2
\right|_{\beta_{m}}$ are the effective diffraction coefficients,
and $\gamma_{m j} = \int_0^h |\psi_b^{(m)} \psi_b^{(j)}|^2 dx$ are
the nonlinear coupling coefficients, where we assume the
normalization $\int_0^h |\psi_b^{(m)}|^2 d x \equiv 1$. It can be
demonstrated~\cite{Kohn:1959-809:PREV} that the diffraction
coefficients $D^{(m)}$ are {\em positive} near the lower gap
edges, and {\em negative} at the upper edges and, therefore, both
{\em bright} and {\em dark} solitons can co-exist in the nonlinear
media with either self-focusing or self-defocusing nonlinearities.
Moreover, it immediately follows from Eqs.~\reqt{nlsc} that
each soliton can support {\em multiple guided modes in other band gaps}, all such modes can be coupled together to form {\em multi-gap vector solitons}. 

The simplified model~\reqt{nlsc} predicts stability of bright
vector solitons~\cite{Pelinovsky:2000-8668:PRE} when all
$\Gamma_{m j}$ have the same sign. However, the applicability of
Eqs.~\reqt{nlsc} is limited to small-amplitude solitons
in the vicinity of the gap edges. Even in this regime, the important effect of {\em the inter-band resonances}, which can lead to the soliton
instabilities~\cite{Sukhorukov:2001-83901:PRL}, is not taken into account by the envelope approximation~\reqt{nlsc}. As the
input power grows, the soliton width decreases and becomes
comparable to the spatial period of the structure. This suggests
that nonlinear properties should depend on the actual profile of
the periodic potential $V(x)$. Modulated periodic structures or
{\em superlattices} can then become an important tool to engineer
both linear and nonlinear properties of the Bloch waves and gap
solitons.

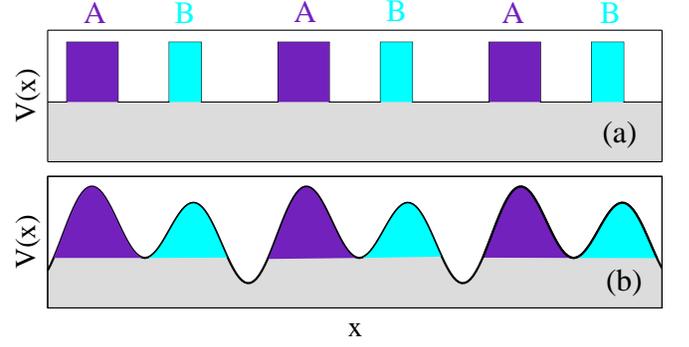
\pict{fig01.eps}{lattice}{Examples of a binary superlattice which can
be created by (a) an array of two types of coupled waveguides, or
(b) an optical superlattice induced by two overlapping mutually
incoherent interference patterns.}

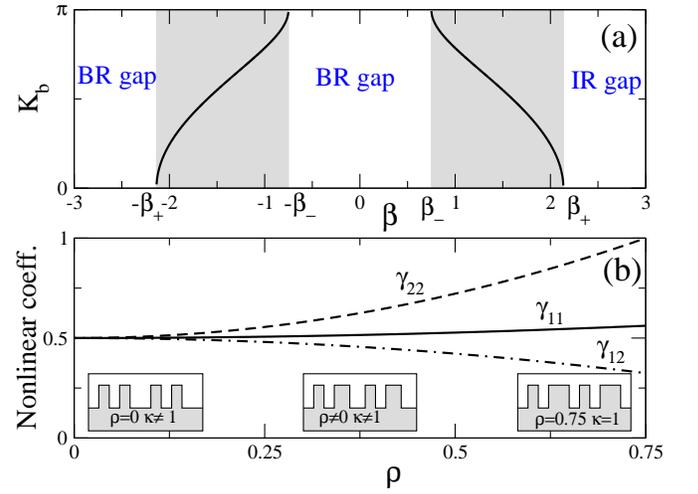
\pict{fig02.eps}{dispers}{ (a)~Characteristic dependence of the
Bloch wave number ($K_b$) on the propagation constant $\beta$.
Gray shadings mark the transmission bands. 
(b)~Dependence of the normalized self- and cross-phase modulation nonlinear coefficients ($\gamma_{m j} = |a^{(m)} a^{(j)}|^2 + |b^{(m)} b^{(j)}|^2$) between the gap edges $\beta_{1} = \beta_+$ and $\beta_{2} = -\beta_-$
vs. the parameter $\rho$. The values of $\beta_{\pm}$ correspond
to the plot (a) by a proper choice of $\kappa$.
Insets show possible symmetries of superlattices
corresponding to different parameter values, but the same linear
dispersion.}

As an example, we consider {\em a binary superlattice} where the
effective periodic potential is composed of two types (A and B) of
separated individual potential wells, $V(x) = \sum_n [V_A(x + n h)
+ V_B(x + n h)]$. Such superlattices can be produced by etching
waveguides on top of a AlGaAs substrate
\cite{Sukhorukov:2002-2112:OL}, or induced dynamically by two
overlapping mutually incoherent interference patterns in a
photorefractive medium \cite{Fleischer:2003-23902:PRL}, see
Figs.~\rpict{lattice}(a,b). In order to analyze the properties of
nonlinear waves of such superlattices, we employ the tight-binding
approximation. This approach allows us to describe correctly the
first two spectral bands, and the Rowland ghost
gap~\cite{Russell:1986-596:PRL} which is defined by a difference
between the A- and B-type lattice cites. We present the total
field as a superposition of the guided modes supported by
individual potential wells, ${\bf \psi}(x,z) = \sum_n [{\bf
A_n}(z) {\bf \psi}_A(x) + {\bf B_n}(z) {\bf \psi}_B(x)]$, where
${\bf A_n}$ and ${\bf B_n}$ are (yet unknown) mode amplitudes.
Finally, we derive a system of coupled discrete equations for the
normalized amplitudes ${\bf a_n}$ and ${\bf b_n}$,
\begin{equation} \leqt{dnls}
   \begin{array}{l} {\displaystyle
      i \frac{d {\bf a}_n}{d z}
      + \rho {\bf a}_n
      + \kappa^{-1} {\bf b}_{n-1} + \kappa {\bf b}_n
      + \chi_a |{\bf a}_n|^2 {\bf a}_n
      = 0 ,
   } \\*[9pt] {\displaystyle
      i \frac{d {\bf b}_n}{d z}
      - \rho {\bf b}_n
      + \kappa {\bf a}_{n} + \kappa^{-1} {\bf a}_{n+1}
      + \chi_b |{\bf b}_n|^2 {\bf b}_n
      = 0 .
   } \end{array}
\end{equation}
Here, the key characteristics of the binary superlattice are
defined by free parameters: $\rho$ is proportional to the the {\em
detuning} between the propagation constants of the A and B-type
guided modes, $\kappa$ characterizes the {\em relative
coupling strength} between the neighboring wells on the right- and
left-hand sides, and $\chi_{a,b}$ are the normalized nonlinear coefficients.

According to Eqs.~\reqt{dnls}, the linear Bloch-wave dispersion is
defined as $K_b = \cos^{-1}( -\eta / 2)$, where $\eta = \kappa^2 +
\kappa^{-2} + \rho^2 - \beta$. The transmission bands correspond
to real $K_b$, and they appear when $\beta_- \le |\beta| \le
\beta_+$, where $\beta_\pm = (\kappa^2 + \kappa^{-2} + \rho^2 \pm
2)^{1/2}$. A characteristic dispersion relation and the
corresponding band-gap structure are presented in
Fig.~\rpict{dispers}(a). The upper gap at $\beta>\beta_+$ is due
to the effect of {\em total internal reflection} (IR), whereas
additional gaps appear due do the resonant {\em Bragg reflection}
(BR).

It is well known that the material dispersion can be completely
compensated by the geometrical dispersion in optical
fibers~\cite{Agrawal:1989:NonlinearFiber}. More recently,
diffraction management was realized in a periodic waveguide array
structure~\cite{Eisenberg:2000-1863:PRL}. The natural open question is {\em
whether it is possible to control nonlinear coupling} between the
gaps by appropriate design of periodic structures. In order to answer
this fundamental question, we study the dependence of the
nonlinear coupling coefficients on the superlattice parameter
$\rho$, while preserving {\em exactly the same linear dispersion} of Bloch-waves by a proper choice of $\kappa$. Our results are
presented in Fig.~\rpict{dispers}(b), and they uncover the
remarkable feature: {\em nonlinear inter-band interaction
coefficients strongly depend on the symmetry of the periodic
structure}, and this relation cannot be fully characterized just
by the linear Bloch-wave dispersion. Thus, by changing the
superlattice parameters it is possible to selectively  {\em
enhance} or {\em suppress} inter-band interaction.

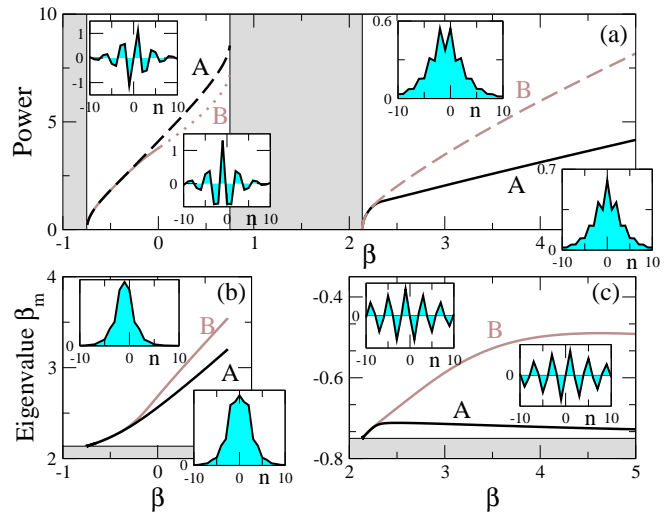
\pict{fig03.eps}{power}{ (a) Power vs. propagation constant for
discrete solitons centered at A (black) and B (gray) lattice
sites. Solid~--- stable, dashed~--- unstable, and dotted~---
oscillatory unstable modes. (b,c)~Eigenvalues of the guided modes
supported by the discrete solitons localized in the complimentary
gap. Insets show characteristic profiles of solitons and their
guided modes defined as $u_{2 n} = a_n$ and $u_{2 n+1} = b_n$. The
array parameters are $\rho = 0.75$, $\kappa = 1$, and
$\chi_{a,b}=1$.}

To be specific, we consider the superlattice with symmetric
inter-site coupling ($\kappa=1$) created in a {\em self-focusing
medium}. Such a lattice can support two fundamental types of
one-component bright solitons centered at either A or B cites, and these solitons exist in both the IR and BR gaps described by the model~\reqt{dnls}. We find that the powers of A- and B-type solitons become significantly different away from the band edges, see Fig.~\rpict{power}(a). 
The solitons of type-B in the IR gap are always unstable with respect to a translational shift (symmetry breaking), however the {\em stability is reversed}
for discrete gap solitons in the first BR gap [left part of 
Fig.~\rpict{power}(a)] where type-A solitons become unstable.
Additionally, the discrete gap solitons become oscillatory
unstable above a critical power due to 
(i)~internal resonance within the gap, first discovered for the fiber Bragg
solitons~\cite{Barashenkov:1998-5117:PRL}, and (ii)~inter-band
resonances, first found for nonlinear defect modes in a layered
medium~\cite{Sukhorukov:2001-83901:PRL}.

The mutual trapping of the modes localized in different gaps and
the physics of multi-gap vector solitons can be understood in
terms of the soliton-induced waveguides~\cite{Ostrovskaya:1998-1268:OL}. Therefore, the effect of discreteness on the inter-gap coupling can be captured by studying the guided modes supported by a scalar soliton in other gaps: the larger is the eigenvalue shift from the band edge, the stronger is the interaction. In Figs.~\rpict{power}(b,c), we plot the
eigenvalues of the guided modes supported by the BR (gap) and IR
solitons and observe {\em two remarkable effects} which cannot be
captured by the simplified envelope approximation~\reqt{nlsc}. First,
the strength of the inter-gap coupling depends strongly on the
soliton symmetry. Indeed, the type B soliton always creates a
stronger effective waveguide, despite the fact that the soliton
power in the BR gap is smaller compared to the type-A solitons.
Second, the nonlinear inter-gap coupling decreases for strongly
localized discrete soliton in the IR regime, as follows from the
non-monotonic dependence of the gap-mode eigenvalues shown in
Fig.~\rpict{power}(c).

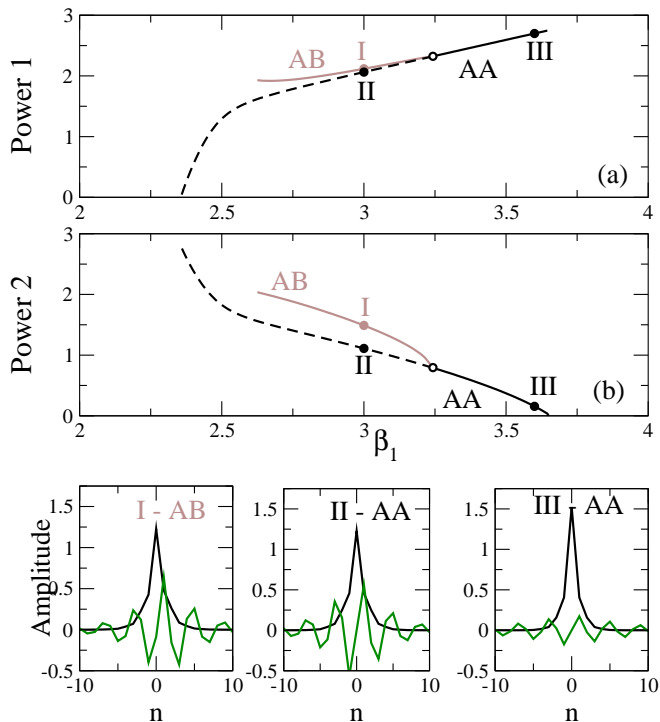
\pict{fig04.eps}{vector}{ (a,b) Powers in the IR (a) and BR (b) components of the multi-gap discrete vector solitons vs. propagation constant $\beta_1$ with $\beta_2 = 0.5-\beta_1 / 3$ for the families with symmetric (AA) and asymmetric (AB)
profiles. Dashed lines mark solutions exhibiting symmetry-breaking
instability. Bottom: Characteristic profiles of the vector
solitons composed of the components localized in two different
gaps: IR~-- unstaggered, and BR~-- staggered; propagation constants correspond to the points marked as I, II, and III in the upper plots. The array parameters match Fig.~\rpict{power}.}

The eigenvalues of the linear guided modes define the point where {\em a multi-gap vector soliton} bifurcate from their scalar counterparts. Initially the amplitude of the guided mode is very small, but it increases away from the bifurcation point, and the mode interacts with the soliton waveguide creating a coupled inter-gap state, see Fig.~\rpict{vector}. In the vicinity of the bifurcation point, the soliton symmetry and stability are defined by the large-amplitude soliton component. For example, the AA-type discrete vector
soliton shown in Fig.~\rpict{vector}(III) is stable because the
powerful A-type mode in the IR gap suppresses instability of the
second component. However, as the power in the second component
grows, the soliton properties change dramatically: (i)~AA state
becomes unstable, and at the same time (ii)~a stable AB-type {\em
asymmetric vector soliton} emerges, see the modes II and I in
Figs.~\rpict{vector}(bottom), respectively. These complex existence and stability properties underline a nontrivial nature of nonlinear inter-gap coupling between the localized components with different symmetries.

We notice that such a simple way to engineer nonlinear coupling in
the superlattices can lead to novel effects in the soliton
collisions. Both coherent interaction of solitons from one band
and vector solitons with the components from different bands can
be controlled by engineering the superlattice parameters, thus
leading to novel features in the soliton switching and steering.
These interesting effects will be discussed elsewhere.

In conclusion, we have studied nonlinear coupling and wave
localization in periodic systems with multi-gap transmission
spectra. We have predicted the existence of novel types of multi-gap vector solitons and studied their stability. Using the example of a binary waveguide array, we have demonstrated the basic concepts of the engineering of nonlinear inter-band interaction in such structures, which in turn determine the key soliton properties.

The authors acknowledge a support of the Australian Research
Council and useful discussions with E.~A. Ostrovskaya.

\end{sloppy}
\end{document}